\documentclass{elsart}
\usepackage{graphicx}
\journal{NIM}
\begin{document}
\begin{frontmatter}

\title{POLAR, a compact detector for Gamma Ray Bursts photon polarization measurements}

\author[ISDC]{N. Produit}\footnote{Corresponding author. Tel +41 22 37 92140.\\
\it{E-mail address:} Nicolas.Produit@obs.unige.ch.},
\author[LIP]{F. Barao},
\author[ISDC]{S. Deluit},
\author[PSI]{W. Hajdas},
\author[Geneva]{C. Leluc},
\author[Geneva]{M. Pohl},
\author[Geneva]{D. Rapin},
\author[LAPP]{J.-P. Vialle},
\author[ISDC]{R. Walter},
\author[PSI]{C. Wigger}  
\address[ISDC]{ISDC, Universit\'e de Gen\`eve, Switzerland}
\address[LIP]{LIP, Lisboa, Portugal}
\address[PSI]{PSI, Villigen, Switzerland}
\address[Geneva]{DPNC, Universit\'e de Gen\`eve, Switzerland}
\address[LAPP]{LAPP/IN2P3/CNRS, Annecy, France}
\begin{abstract}
The design and the simulated performances of a compact detector dedicated to
the measurement of GRB photon polarization is presented. Such a detector would
permit to answer the question ``are most of the GRB strongly polarized?''
in a mission of one year in space.
\end{abstract}

\end{frontmatter}

\section{Introduction}
Discovered 35 years ago, Gamma Ray Bursts (GRB) remain a challenging issue and
one of the most interesting topics in astrophysics. The genuine character of the GRB
sources and the energy release mechanisms still remain a mystery and there are several
competing theoretical models and mechanisms trying to explain it.

The launch of several high energy observatories with powerful instruments on board 
(XMM, Integral, RHESSI, BeppoSax) and of advanced GRB detectors (CGRO, HETE2) made possible a
number of new discoveries and in parallel noteworthy theoretical progresses.
However the instruments advances were mainly on effective area, energy and angular resolution,
while very little progress was achieved on polarization measurement, partly due to the experimental difficulty
of measuring polarization. Today, photon polarization turns out to be one of the ultimate observables
required to make a clear distinction between different theoretical models in order to fully understand
the GRB nature. To date however, the polarization was explored only in a few afterglow
measurements in visible light.

The recent report\cite{Coburn} of a high linear polarization (80$\pm$20\%) in the prompt emission of
GRB021206, together with its subsequent revisions as too preliminary
(if not mistaken) \cite{Rutledge} \cite{Wigger}, demonstrated that polarimetric measurements
are difficult and need dedicated instruments. Such measurements are a must in the era of missions
like SWIFT\cite{SWIFT} or the forthcoming GLAST\cite{GLAST} experiment. Both SWIFT and GLAST 
can promptly determine the exact
GRB coordinates as well as the lightcurve and spectral characteristics. For such bursts the the
polarimetric detectors will bring simultaneously additional
unique information about the magnetic field structure and dynamics during the prompt
gamma ray emission. We propose here a dedicated instrument with high sensitivity for polarization measurements.

\section{GRB's and polarization}

Many theoretical models of GRB's give polarization predictions for both, prompt and afterglow emissions [6-12].
The commonly accepted fireball model requires a fine tuning of the magnetic field structures in order 
to provide high values of polarization. In addition, it predicts similar 
polarization values for prompt and afterglow emissions. Other models, like 
the electromagnetic and the cannonball, allow for high polarization of photons 
in a more straightforward manner. As a matter of fact, predictions are not unique but 
rather give a range of polarization levels as, like in the cannonball model, the
value of the polarization depends not only on internal parameters but also on the observer's viewing angle.
Unambiguous distinction between models thus demands 
multiple polarization measurements from a large number of bursts, which in turn implies a dedicated polarization detector.

It is extremely difficult to cover a wide energy range while getting the
high detection sensitivity needed for the GRB studies within a single spacecraft instrumentation.
Even the recently launched top-of-the-art GRB observatory SWIFT has a photon
detection upper energy limit of about 150~keV, and it does not carry a gamma
ray polarimeter on board. At present, the GRB measurement approach 
is to conduct a number of multi-wavelength observations done both simultaneously and 
in a follow-up manner with multiple detector systems usually installed on different 
satellites linked together. Such a global network (GCN \cite {GCN}) synchronizes various 
observations and provides in quasi real time GRB detection alerts to many 
follow-up instruments like optical telescopes. Though, despite of many attempts supported 
by wide theoretical efforts, polarization data remain very scarce and measurements 
very difficult. There are only few results reporting polarization measurements in 
the GRB afterglow data \cite{afterglow}. At typical gamma ray energies, even novel instruments with designed 
polarimetric capabilities like e.g. RHESSI\cite{RHESSI} with its passive 
Beryllium scatterer face problems with systematic effects. In coincidence mode, 
the RHESSI polarimeter has very small active area and suffers from either poor statistics 
or high background levels or both\cite{Wigger}. Other satellites like Integral\cite{integral} have their 
instruments not optimized for polarization measurements. Therefore, their efficiency,
analyzing power or background suppression capability is too low for any meaningful long-term observations.

Bearing in mind the outstanding importance of polarization measurements, we propose a 
GRB polarimeter based on Compton scattering and on well known detection technologies.
The design is optimized for an energy range from a few keV up to several hundred keV and a wide viewing angle. The instrument is 
characterized by a large area and high analyzing power. It utilizes low Z,
fast scintillation square detector bars arranged in a large array. Full description 
of the polarimeter and first results from its extensive modeling are presented below.

\section{Detector requirements}

\begin{figure}
\begin{center}
\includegraphics[width=0.4\linewidth]{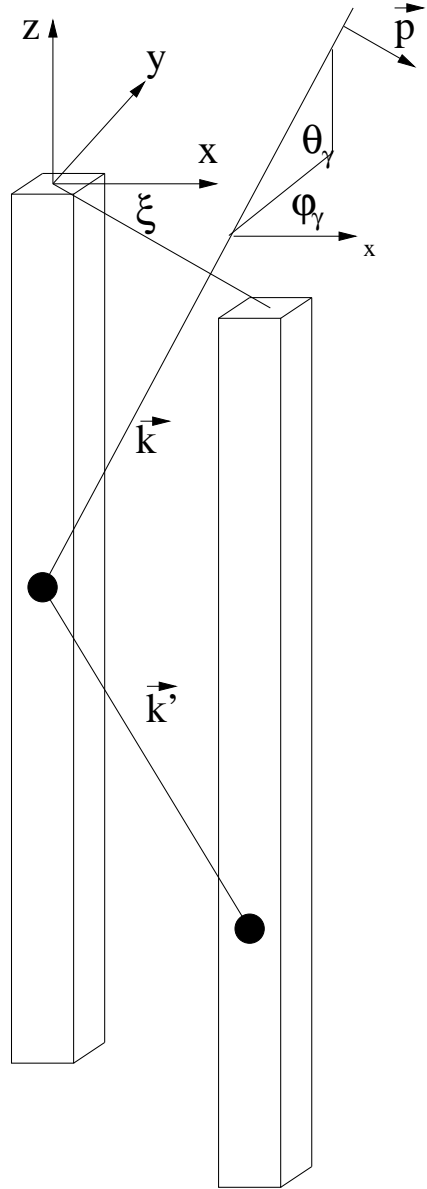}
\end{center}
\caption{Geometry of the large angle Compton scattering. The two bars
where interaction occur are shown. $\theta_\gamma$ and $\phi_\gamma$ are the
entrance angle of the photon relative to a detector fixed coordinate system.
$\xi$ is the measured azimuthal direction that correlate with polarization 
direction $\protect\overrightarrow{p}$.
}
\label{fig:comptongeom}
\end{figure}

The method for measuring gamma ray polarization depends on their
energy. In the energy range of 10 keV to 1 MeV, Compton scattering
can be used thanks to its large cross section and its sensitivity to
polarization (see geometry on Fig.~\ref{fig:comptongeom}).
Small angle scatterings dominate the total cross section, 
but they basically conserve the direction of polarization of the incoming photon, thus they do not affect the measurement. 
In large angle scatterings, the azimuthal ($\xi$) angular distribution is strongly modulated around the photon
polarization direction. It can be reconstructed by observing the recoil electron from the Compton 
scattering and then observing the 
scattered photon by a subsequent process depositing sufficient energy
(second Compton scattering or photoelectric effect).
In addition, the cross section is symmetric under $\xi \rightarrow \pi-\xi$, such
that it is not necessary to know the order in time of the two observations. Monte Carlo simulation clearly
indicate that it suffices to
isolate the two most energetic processes. The line connecting them is well correlated
with the outgoing azimuthal photon direction $\xi$.     

In the design of the detector, the following key features must be taken into account :
\begin{itemize}
\item at X-ray and soft gamma ray energies the radiation and the absorption lengths are short enough such that a rather compact detector using low Z materials is conceivable.
\item The detector must tolerate multiple small angle scatters, favor
large angle Compton scatterings, and offer a good resolution for the $\xi$ angle. 
\item Charged particles, entering with an overwhelmingly high rate, must not leave a signal 
that could be confused with the desired photon signal.
\item Photons from the spacecraft itself, from diffuse radiation and from usual 
sources form together an irreducible background. Signal-to-background ratio thus
call for a fast detector response with good time resolution.
\end{itemize}

\begin{figure}
\begin{center}
\includegraphics[width=1.0\linewidth]{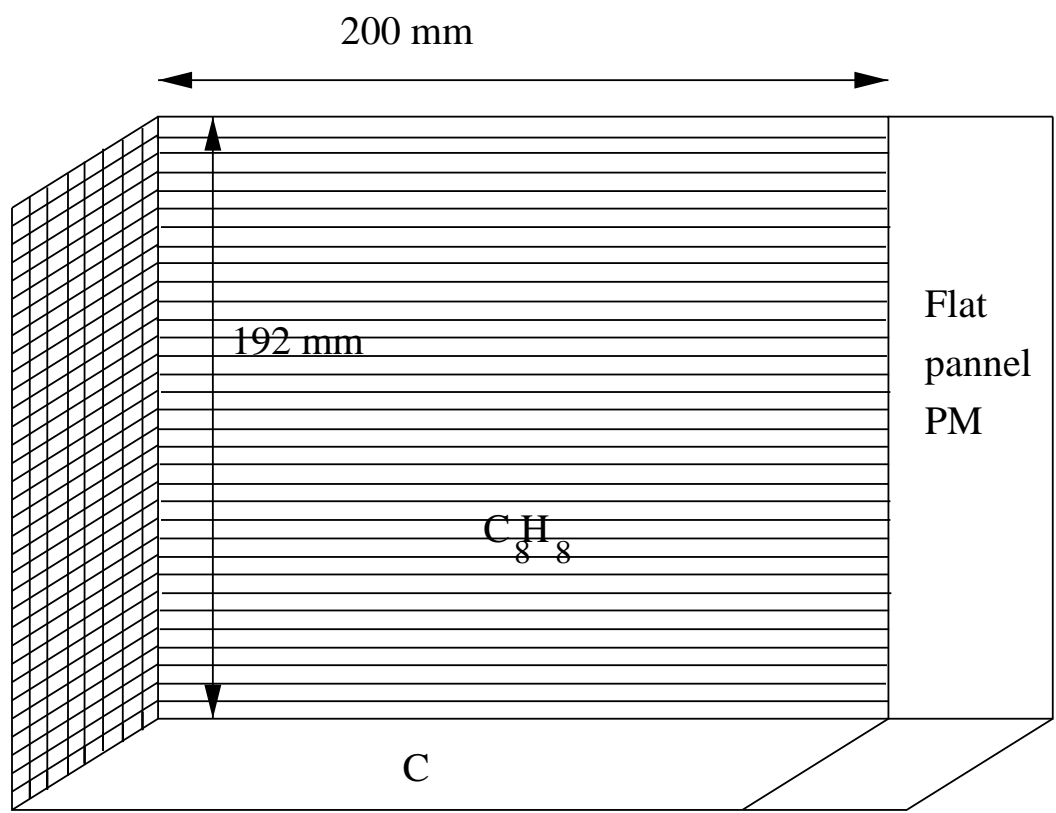}
\end{center}
\caption{Schematic view of the POLAR detector.}
\label{fig:polardetector}
\end{figure}

These considerations lead us to a design (See Fig.~\ref{fig:polardetector})
consisting of an homogeneous and symmetric 
active target, made of not very high $Z$ material, like plastic scintillator. 
A target segmentation of a few $mm^2$ was dictated by the range of recoil electrons 
in the given photon energy range as well as the required angular resolution.
This segmentation matches well the anode dimensions of modern metal mesh
multi-anode photo multipliers (PM), which have already found
wide-spread use in space missions 
and provide the performances required.

\section{The POLAR Detector}
The detector is an active target of outer dimensions $192\times192\times200 mm^3$. The target
material is plastic scintillator, i.e.~doped polystyrene, chemically and mechanically 
stable and supporting high total radiation doses with little degradation. The target
is segmented into 2304 elements of dimension $4\times4\times200 mm^3$, with their long 
axis facing the preferred photon entry direction. The target elements are optically insulated 
from each other by applying a thin double layer of reflective/diffusive paint.

The front face of the target is shielded by a 
1~mm carbon fiber shield to absorb very low energy photons and charged particles. 
Likewise the sides are passively shielded by a 1~mm thick carbon fiber,
which also provides a mechanical enclosure. On the back, a flat panel metal mesh 
photo detector is directly coupled to the scintillator elements. With the 
associated power distribution and front-end electronics, the photo detector is thick
enough to shield the detector from photons entering from the back.  

\section{Trigger strategy}
\begin{figure}
\begin{center}
\includegraphics[width=1.0\linewidth]{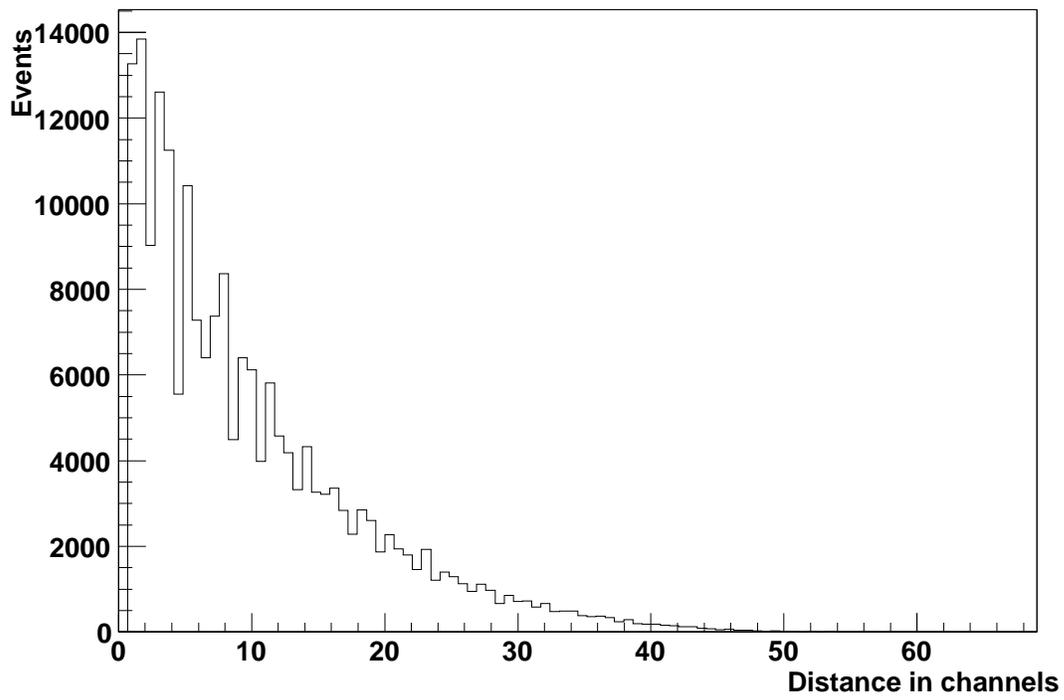}
\end{center}
\caption{Distance in scintillator elements width between the two largest energy deposition.}
\label{fig:distance}
\end{figure}

The signal for a Compton scattering consists of at least two energy depositions, from the 
recoil electron in a large angle scatter and from a subsequent photon interaction via
additional Compton scattering or photoelectric effect. The trigger logic 
thus requires at least two coincident channels each with at least 5 keV energy deposition.
The two elements with the highest energy deposits define the geometry of the event.
The photon entrance angles are assumed to be known, they are derived from the knowledge of the spacecraft
attitude and from the GRB sky coordinates given by another spacecraft or the GCN. 
The azimuthal angle of the straight line connecting the two highest energy hits around 
the entrance axis thus defines the $\xi$ angle sensitive to polarization..  
There is no requirement of maximum distance between the two highest
energy deposits,
but a criterion could be defined without large acceptance losses (see Fig.~\ref{fig:distance}).

\begin{figure}
\begin{center}
\includegraphics[width=1.0\linewidth]{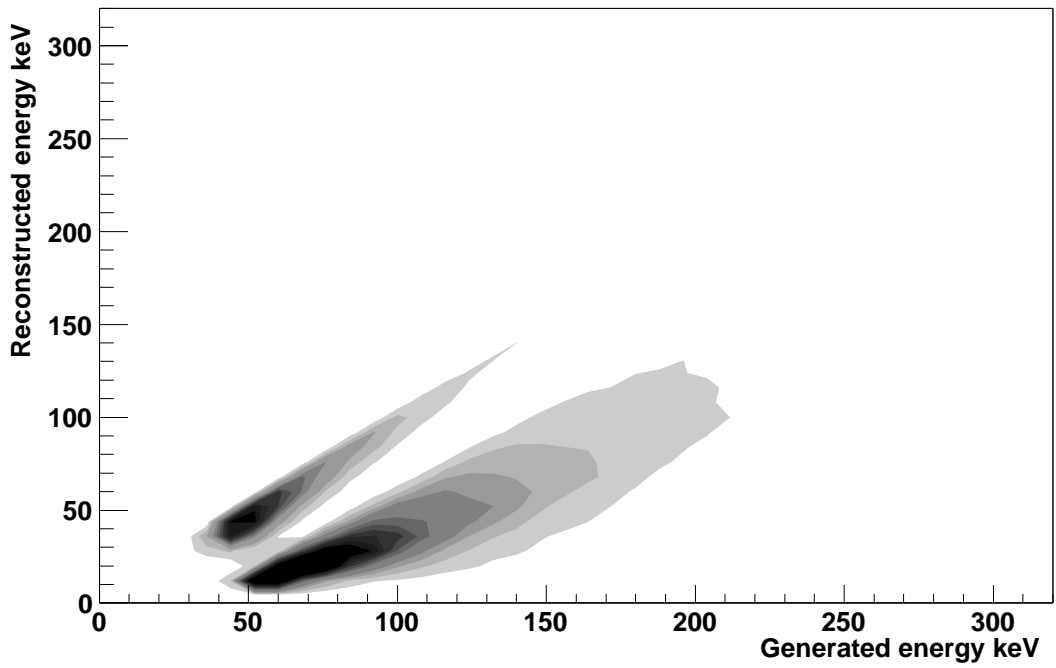}
\end{center}
\caption{Primary energy (abscissa) vs reconstructed deposited energy (ordinate).
Diagonal events are fully contained. For most of the events (the lower population),
some energy escapes the detector.}
\label{fig:erecegen}
\end{figure}

A charged cosmic ray traversing the detector will hit several plastic bars and deposit 
in each of them at least 800 keV. To get rid of this
background the trigger requires
 that the total energy deposition in the target is less than 300 keV. This cut has little
effect on the signal photons (see Fig.~\ref{fig:erecegen}),
since the photon energy spectrum of GRB's is steep, and the detectability of
polarization favors low energy photons. 

Using a typical light yeld of 1 photon/90 eV, 5 keV primary
energy deposition corresponds to about 55 primary scintillation photons.
A conservative light collection efficiency of 10\% and a
typical quantum efficiency of 0.2 shows that
an electronics sensitive to single photoelectrons will reach this threshold.
A lower threshold is not desirable as the background increase faster then
the signal for low energies.

\section{Physics Simulation}
The interactions of signal and background photons with the detector has been simulated using 
GEANT 4\cite{geant}. 
The target is simulated as a solid block of $C_8H_8$ material with appropriate density and
radiation length. The high voltage divider and electronics material on the back end of the target
was grossly approximated as a 10~cm deep block of aluminum with reduced density 
(0.9~$gcm^{-3}$, i.e. 1/3 of pure aluminum). The front and side shield are taken as 1 mm of carbon of density 
2.265 $gcm^{-3}$. The total weight of the detector simulated is then 11.4~kg.

\begin{figure}
\begin{center}
\includegraphics[width=1.0\linewidth]{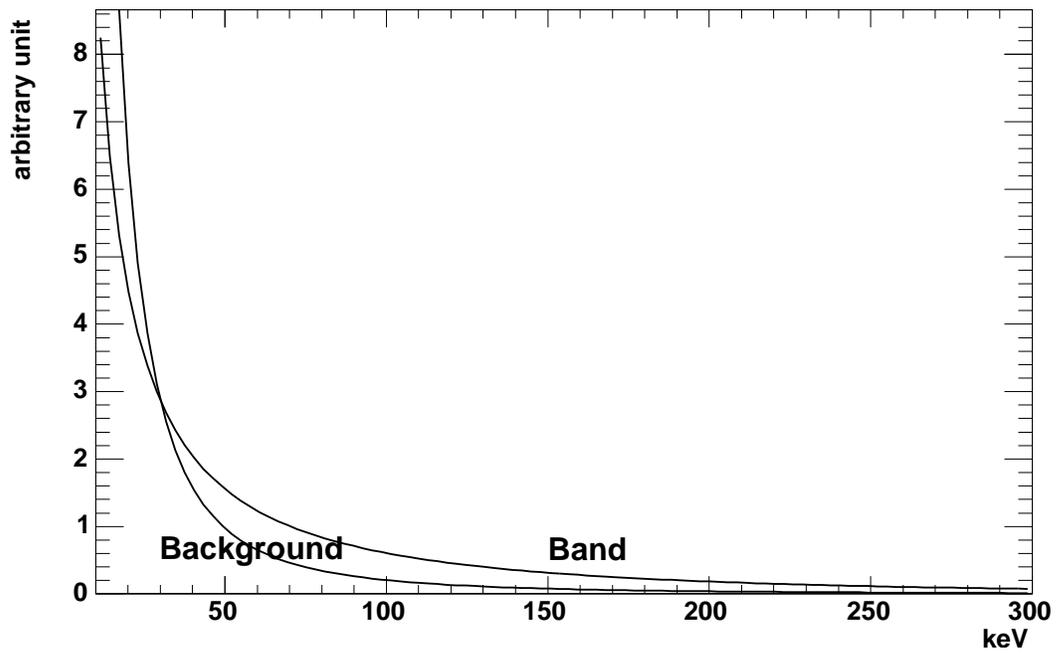}
\end{center}
\caption{Band function used to simulated the photon spectrum emitted by GRB's. 
For comparison the distribution of diffuse background is plotted scaled to the same integral flux 
in the energy range between 10 keV and 300 keV.}
\label{fig:spectrum}
\end{figure}

Incoming photons have been generated uniformly over the front surface,
with entrance angles of $0^\circ \leq \theta_\gamma < 70^\circ$ 
and $\phi_\gamma = 0^\circ$ or $45^\circ$.
Photons generated are fully polarized with a polarization vector $\phi_0=0$ parallel to
one of the detector faces, but is has been checked that the results of the study remains valid for other directions.
The energy is sampled from a distribution based on a band model~\cite{band}
with parameters $\alpha=-1$, $\beta=-2.5$ and $ E_{peak}=200$ keV, between 10~keV and 300~keV.  
The corresponding spectrum is shown in Fig.~\ref{fig:spectrum}, compared to a typical 
background distribution (see section~\ref{sec:background}).

The simulation of physics processes includes the polarization dependence of Compton scattering, and  takes into account the electromagnetic processes which are all implemented in GEANT 4, even those at very low energy. As expected, the photon interactions are dominated in number by low energy
Compton scattering with small energy transfer to the electron. Whenever a large angle
scattering occurs, a clear modulation of the azimuthal angle is observed.
On average, 14\% of the events fulfill the double energy deposition coincidence required by the trigger.

Fig.~\ref{fig:distance} shows the distance between the two largest energy depositions in units
of element width. The leveling off of the distribution justifies the target dimensions. 
Since there are few entries at large distances, a maximum distance requirement can be introduced
if required by combinatorics, at the cost of a small efficiency loss.

Fig.~\ref{fig:erecegen} shows the reconstructed total energy in the target versus the energy of 
the incoming photon. 
It is clearly seen that for a fraction of the events the full energy is collected while for others
only a large angle Compton scattering deposition occurs, the outgoing photon escaping from the detector.

\begin{figure}
\begin{center}
\includegraphics[height=0.5\textwidth]{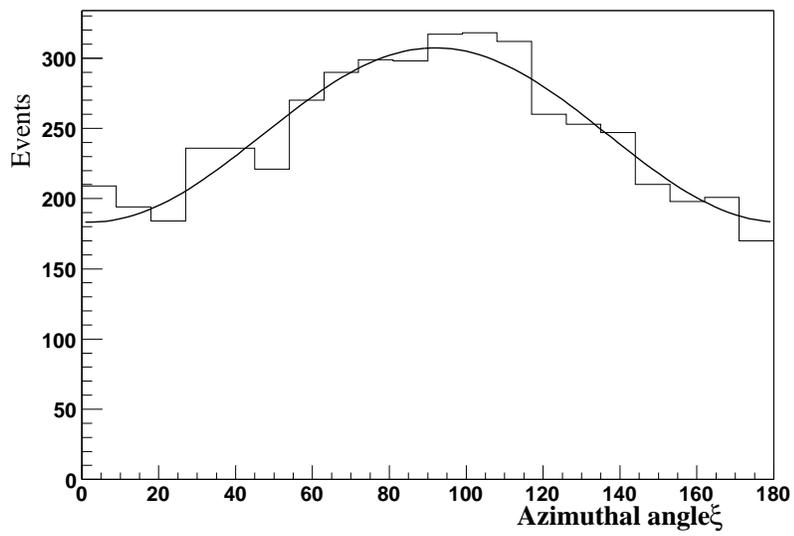}
\end{center}
\caption{Example of a modulation curve measurement for 
a very intense signal.}
\label{fig:grbex}
\end{figure}

As an example,~Fig.~\ref{fig:grbex} shows the reconstructed 
azimuthal angle distribution for 
a very intense GRB.
For a polarization orientation along $\xi_0 = 0^\circ$, the maximum 
signal rate should be at $90^\circ$, which is indeed observed. The 
observed rate as a function 
of $\xi$ during a fixed time interval of the burst is parameterized as 
$\frac{dN}{d\xi} = A (1+ B \cos{2(\xi - \xi_0)})$.
The modulation B indicates the degree of polarization, while the phase $\xi_0-\pi/2$ indicates
the polarization orientation.

\begin{figure}
\begin{center}
\includegraphics[height=0.5\linewidth]{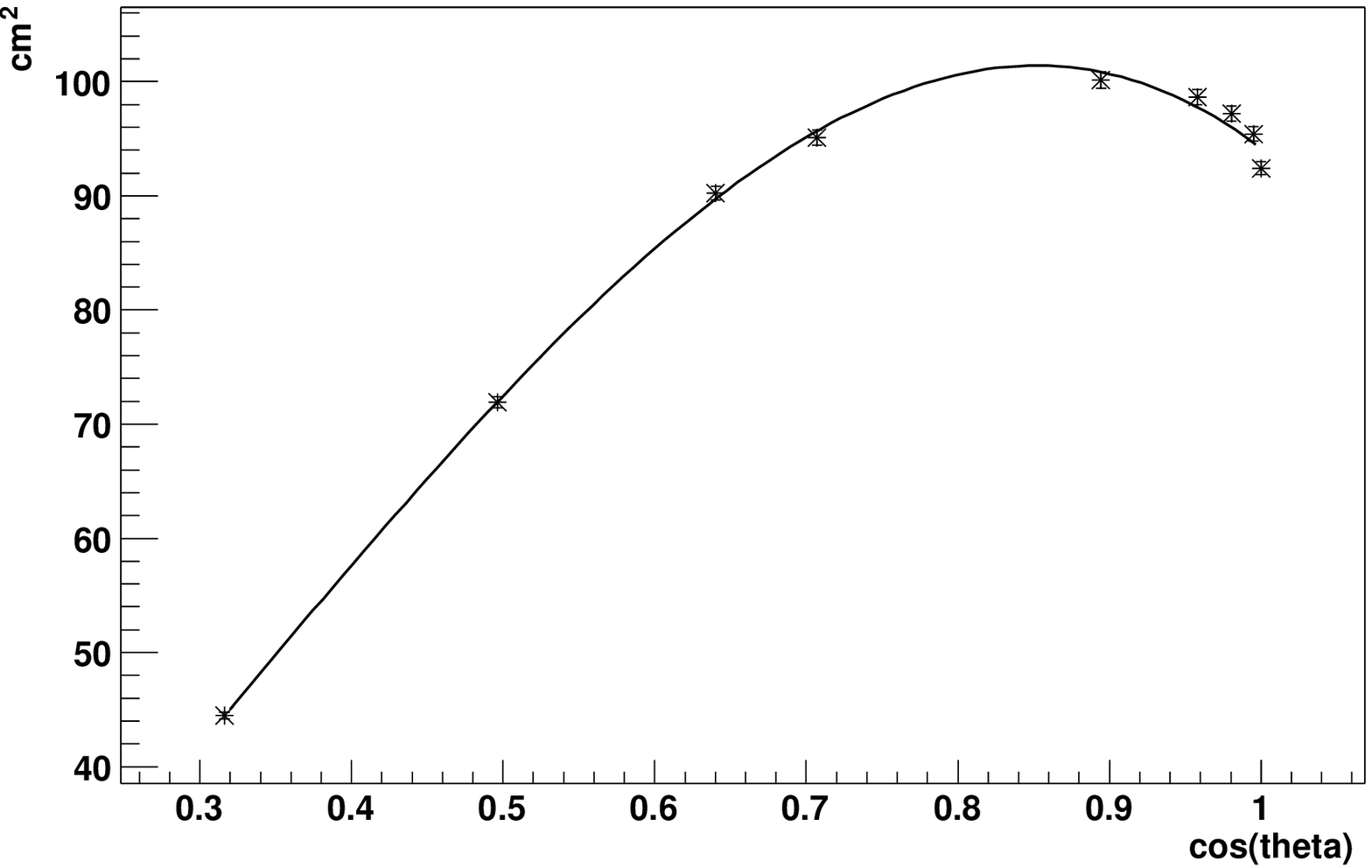}
\includegraphics[height=0.5\linewidth]{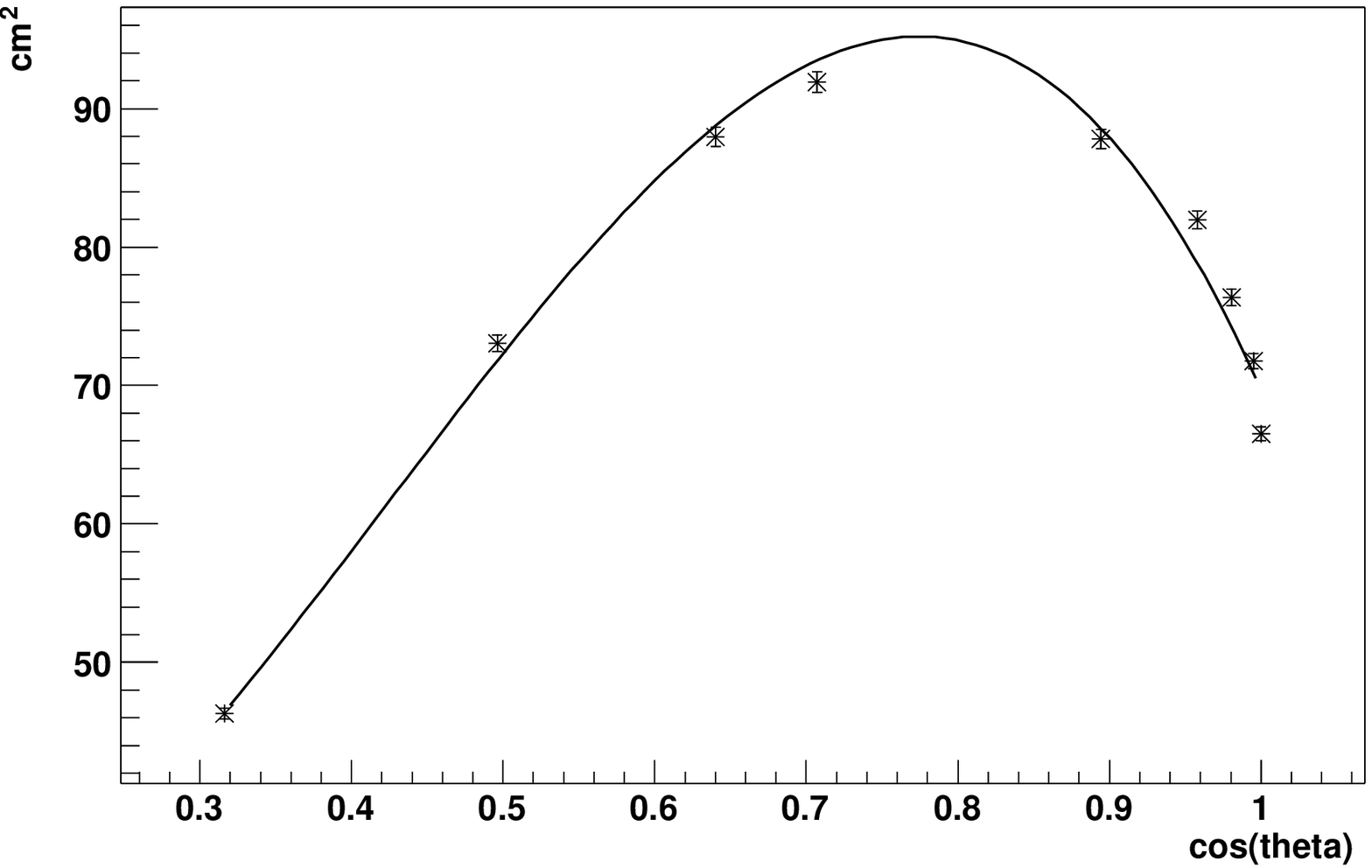}
\end{center}
\caption{Effective surface of the detector in $cm^2$ for photons with energies between
10 and 300 keV as a function of $\cos{\theta_\gamma}$ for $\phi_\gamma = 0$ and $\phi_\gamma = 45^\circ$.}
\label{fig:acceptance}
\end{figure}

The effective area, given by the geometrical surface multiplied by the photon detection probability,
 varies from 100~$cm^2$ to 40~$cm^2$, depending on the photon entrance angle. 
As shown in Fig.~\ref{fig:acceptance}, the acceptance varies slowly over about 1.3~$\pi sr$.
Within this fiducial solid angle the average area is 80~$cm^2$.

\begin{figure}
\begin{center}
\includegraphics[height=0.5\linewidth]{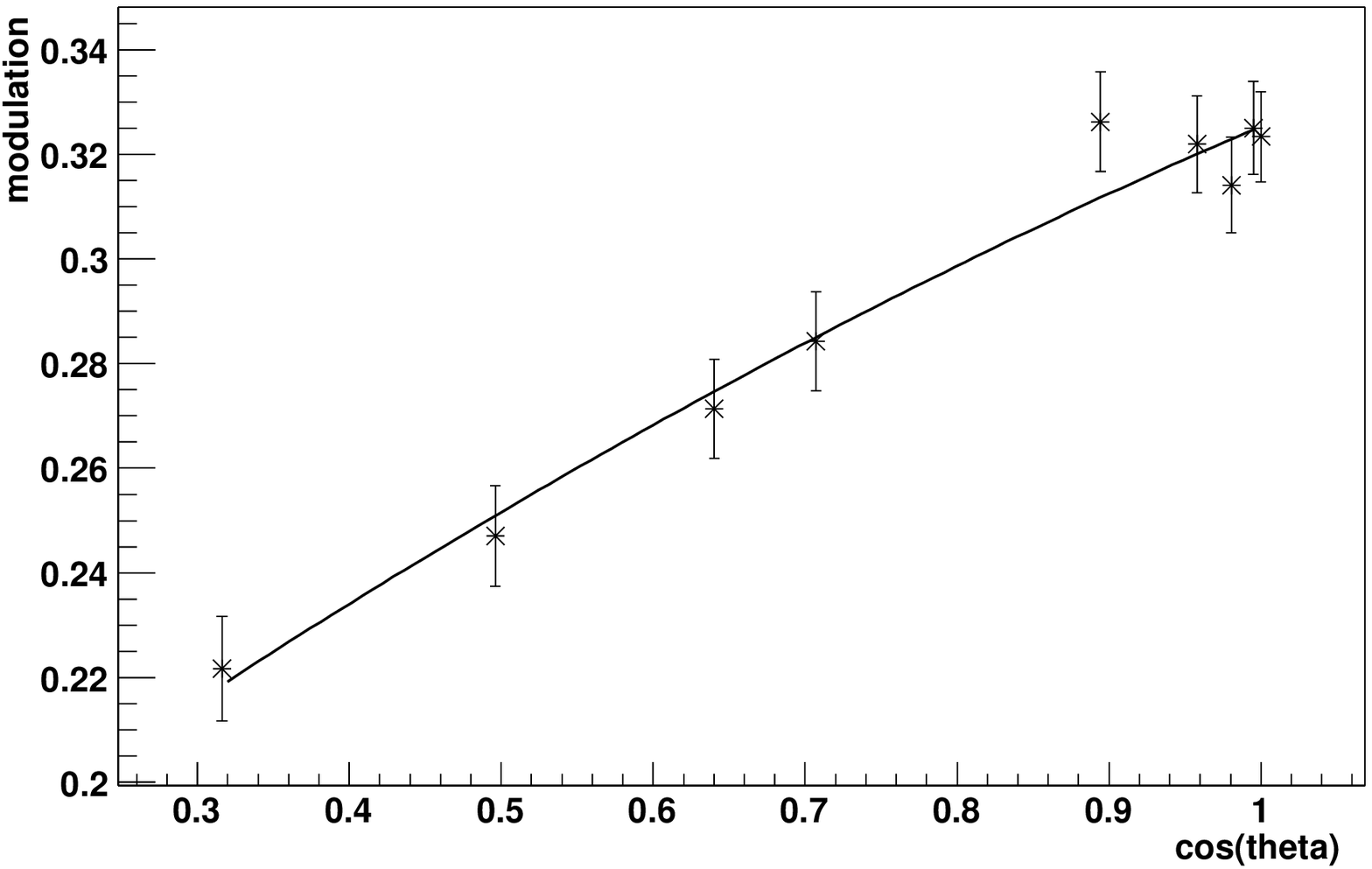}
\includegraphics[height=0.5\linewidth]{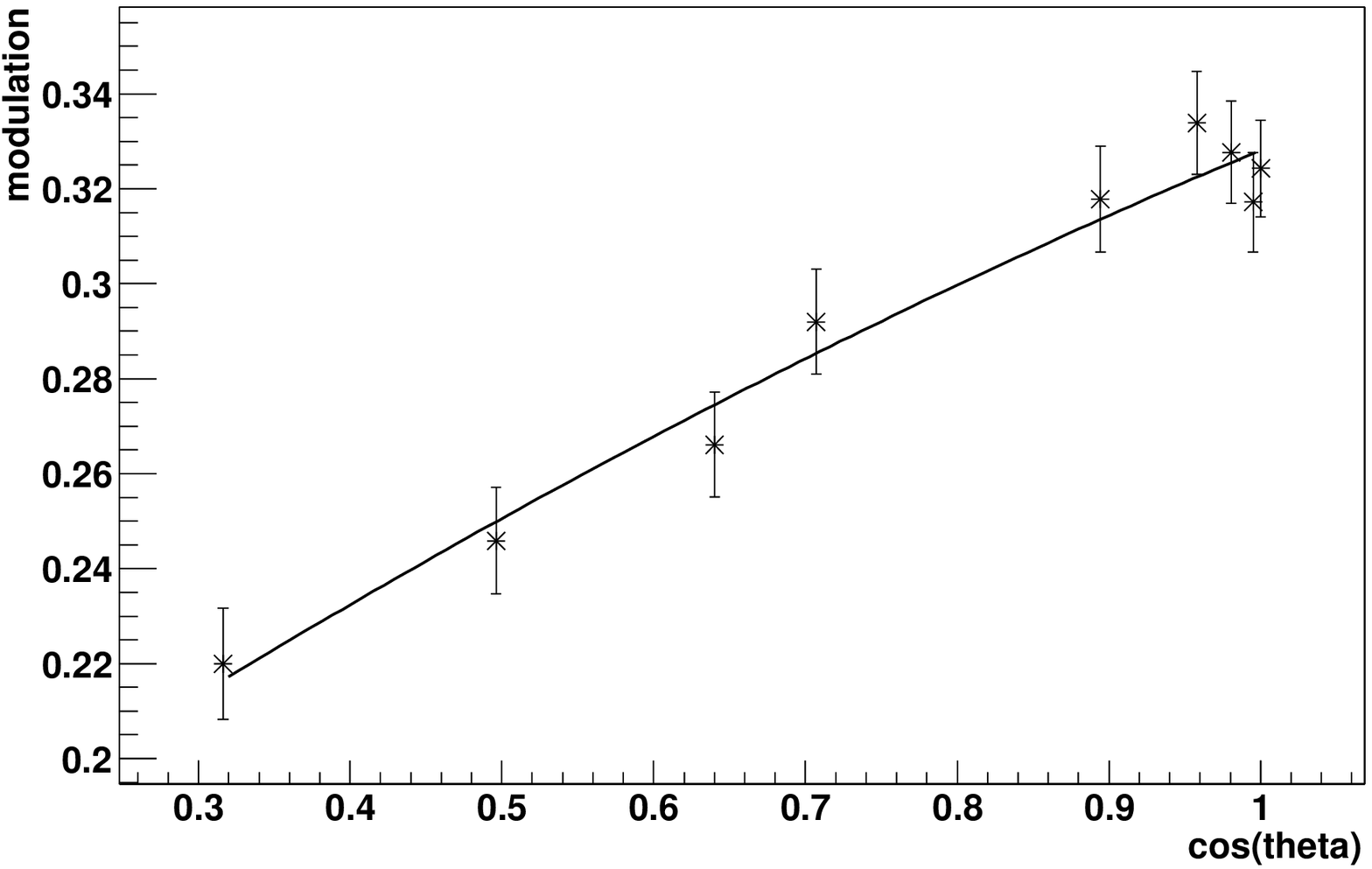}
\end{center}
\caption{Modulation of the azimuthal angle distribution as a function of the
photon polar angle
$\cos(\theta_\gamma)$ for $\phi_\gamma = 0$ and $\phi_\gamma = 45^\circ$. 
For typical entrance angles, a mean modulation of about 0.28 is expected.}
\label{fig:modulation}
\end{figure}

Depending on photon energy and impinging angle, the modulation varies from 0.34 to 0.2, as shown
in Fig.~\ref{fig:modulation}. In further estimates, an average modulation of 0.28 will be used.

\begin{figure}
\begin{center}
\includegraphics[height=0.6\textwidth]{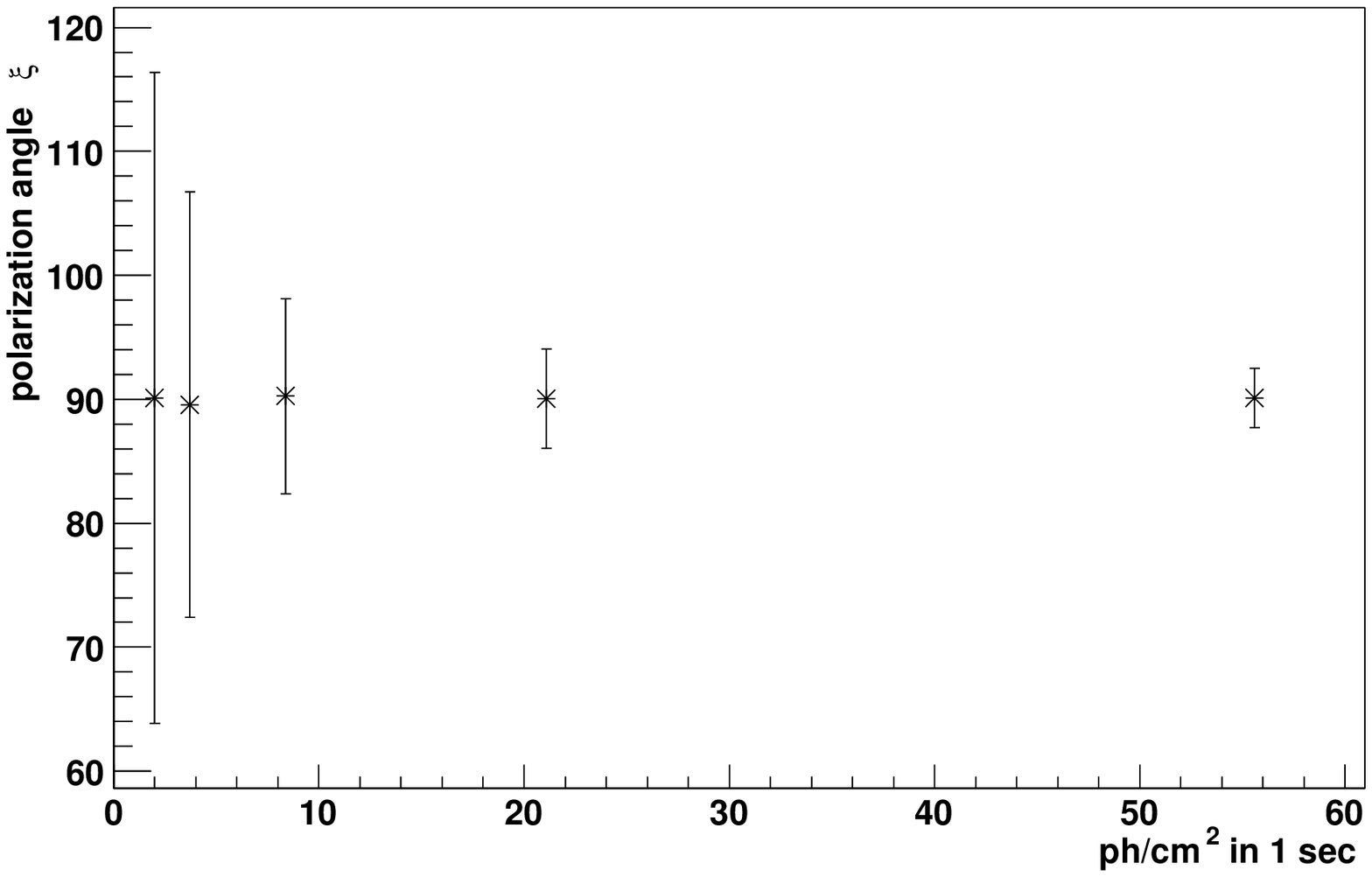}
\includegraphics[height=0.5\textwidth]{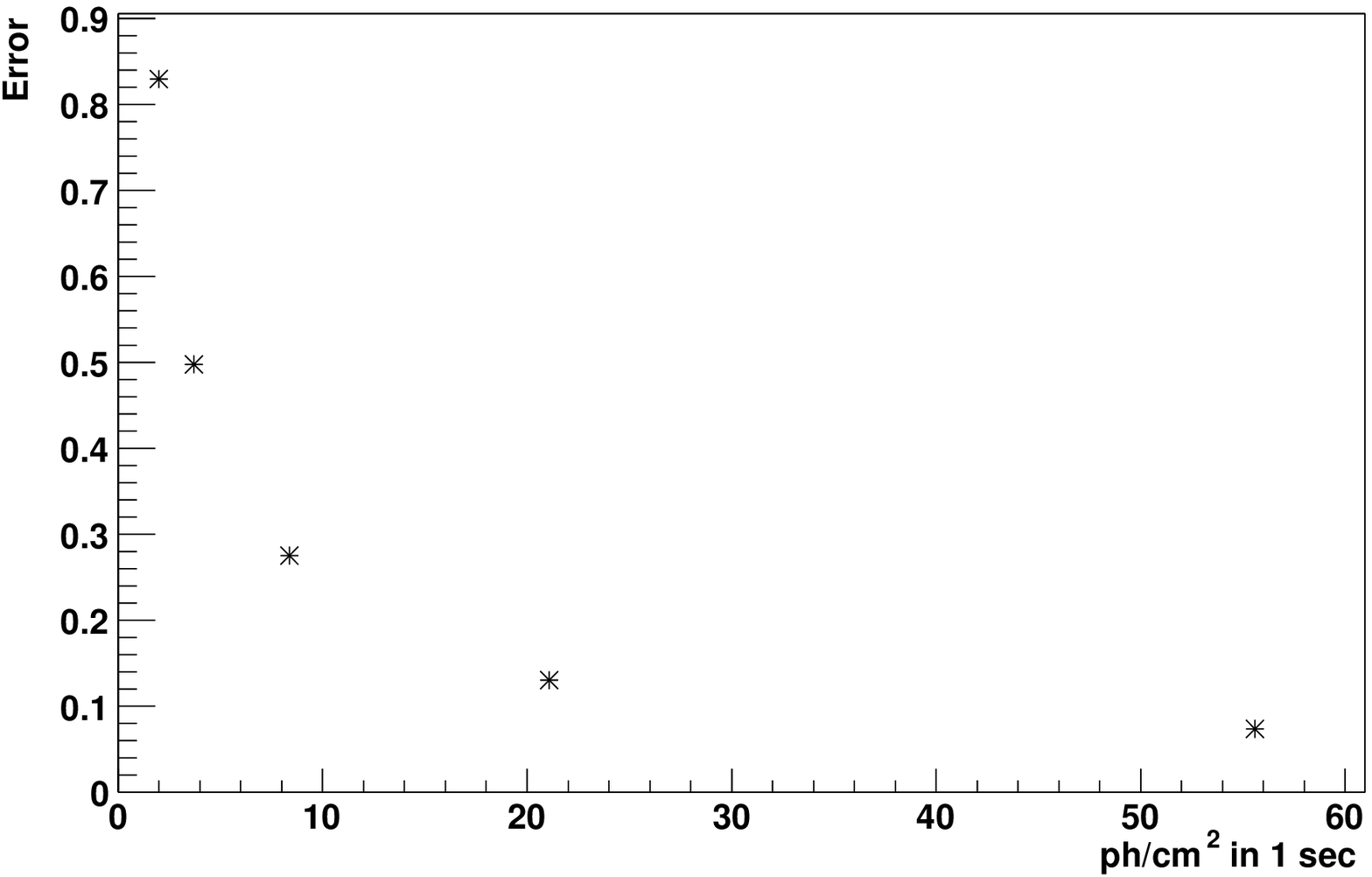}
\end{center}
\caption{Polarization orientation $\xi_0$ and reconstructed
error on the degree of polarization as a function of
the GRB flux during a 1s interval.}
\label{fig:grbmod}
\end{figure}

\begin{figure}
\begin{center}
\includegraphics[width=1.0\linewidth]{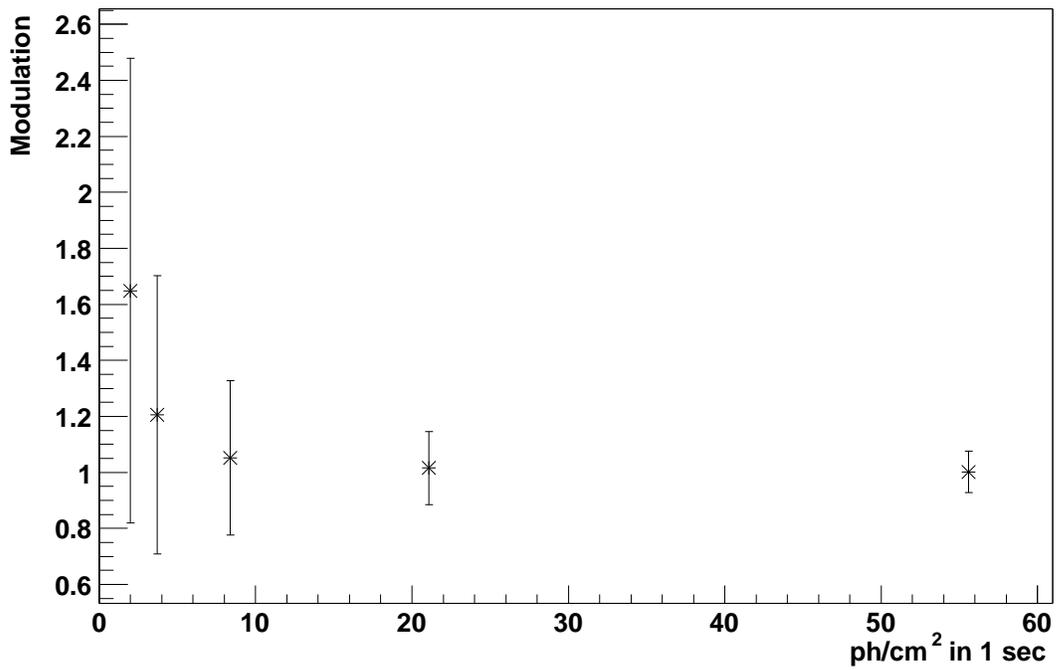}
\end{center}
\caption{Mean value and r.m.s. of the reconstructed degree of polarization, extracted from 
a large sample of simulated GRB. At measurable degrees of polarization, the measurement is 
unbiased and exhaustive.} 
\label{fig:bias}
\end{figure}

Fig.~\ref{fig:grbmod} shows the measured
polarization orientation and the error on the degree of polarization in
function of the intensity of the GRB.
If the flux is not sufficient, no measurement of polarization is possible but there is no 
systematic bias in the measurement as seen on Fig.~\ref{fig:bias}.

\section{Background estimate}\label{sec:background}

Table~1 summarizes the expected trigger rate from the different processes
of signal and background and their impact on the dead time of the detector.
When a signal $> 5 keV$
induce a signal in one or many PM channels, those channels go dead for 100 nsec (PM hits).
When a coincidence
of at least two PM with proper energy range hits occur, the detector is read out.

High energy charged particle background is eliminated completely by the upper energy cut in the trigger. They induce
PM hits but no readout and no dead time.
Induced beta and alpha radioactivity is also not likely to pass the trigger selection except if they take place
exactly between two bars. If they turn out to be a problem we can eliminate them with a cut excluding adjacent
bars with a low penalty as can be seen on Fig.~\ref{fig:distance}.
Low energy electrons from the radiation belts are a more serious nuisance. Electron under
500~$keV$ are stopped by the 1~$mm$ carbon shield. Electrons between 50~$keV$ and 1.5~$MeV$ have a $10^{-3}$ 
probability to look like an real event (d in the table). They are so numerous
that a random coincidence of two of them is likely (noted r in the table).
An orbit like the ISS is spending 20\% of the time 
in the polar horns and in the South Atlantic anomaly.
When in those region the detector will receive
$10^3 cm^{-2}s^{-1}$ electrons in this range. If they turn out to be a real nuisance, the
detector will be switched off when crossing those regions.
All other electrons are vetoed by the trigger.

Protons from the belt are no problems. Protons $<13 MeV$ don't enter the sensitive
volume. Proton $> 13 MeV$ are rejected
by the trigger. In the polar horns and South Atlantic anomaly there is about 10 protons $cm^{-2}s^{-1}$ of energy greater than 13~$MeV$.

There are four sources of background of photons: i) induced radioactivity in the 
scintillator  ii)induced radioactivity in the supporting spacecraft iii) non-GRB point sources
and iv) the diffuse $\gamma$ ray flux.
Induced $\gamma$ rays flux depend very much on the orbit and on the design of the 
spacecraft. As a rough estimate we shall take these numbers from the Integral mission. 
In the ISGRI instrument of Integral, there is a background of 0.02 photons~$cm^{-2} s^{-1} sr^{-1}$ in a similar 
energy range.

The diffuse photon background from the sky has been parameterized~\cite{gammasky} as 
$\log{f(E)} = a + b \log{E} + c \log{E} \log{E}$ photons $cm^{-2} s^{-1} sr^{-1} keV^{-1}$, with 
$a=0.940059$, $b=-1.28089$, $c=-0.262414$ and $E$ in $keV$.
This parameterization predicts 2.46~photons $cm^{-2} sr^{-1} s^{-1}$ with $E_\gamma > 10 keV$.
A source like the Crab, with a flux of $9.7 (E/1\mbox{keV})^{-2.1}$ photons $cm^{-2} s^{-1} keV^{-1}$ 
is contributing 0.7 photons $cm^{-2}s^{-1}$  in the energy range from 10 to 300 keV.

\subsection{Sensitivity to background photons}

\begin{figure}
\begin{center}
\includegraphics[width=1.0\linewidth]{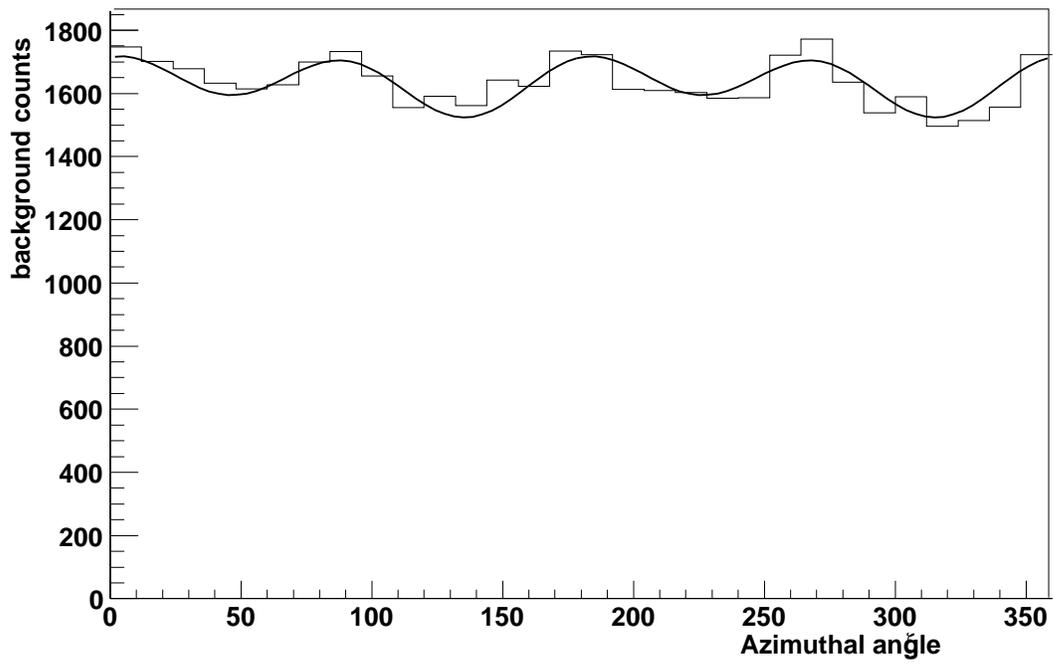}
\end{center}
\caption{Azimuthal angle distribution of background photons. The detector geometry induces a 
modulation with a $\pi/2$ periodicity with respect to one of its axes, which is easy to 
model and subtract.}
\label{fig:background}
\end{figure}

Photons distributed like the diffuse background between 10 keV and 300 keV,
coming from the $2\pi sr $ of sky above the detector have been simulated.
The acceptance for these photons has been found to be 175.2~$cm^2$sr.
The reconstructed polarization angle has a structure incompatible with the signal. 
As the detector is a cube we expect to see a modulation with period $\pi/2$.
Therefore the background has been fitted with a
function $\frac{dN}{d\xi} = A (1 + B \cos{4(\xi-\xi_0)}+C\cos{2(\xi-\xi_1)})$ as
it is shown in Fig.~\ref{fig:background}.

Using the efficiency found for photons distributed according to such an energy distribution, 
a background rate of 430 counts $s^{-1}$ is predicted in the target.

\section{Minimum detectable polarization levels}
A toy Monte Carlo calculation was used to estimate the minimum polarization detectable.
Neglecting the modulation induced by the target geometry, a flat background 
superimposed with events distributed like a 100\% polarized GRB has been made.
Statistical 
fluctuations of the background and of the signal rate has been properly taken into account. A maximum likelihood 
fit to the distribution has been used for extracting the modulation as 
well as the phase angle. 

The minimum detectable polarization level can be defined as 
$MDP=\frac{n_{\sigma}}{\mu S}\sqrt{\frac{S+B}{T}}$ with $n_{\sigma}$ is the number of $\sigma$
we want for our signal, S the signal rate, B the background rate an $\mu$ the modulation for a 100\%
polarized signal. With this formula using $n_{\sigma}=3, \mu=0.28 B=430 photon s^{-1}$ 
and an effective area of 80~$cm^2$, 
a MDP of 100\% is reached with a peak flux of 3.6 photons $cm^{-2}s^{-1}$ in a 1 sec
measurement.

\subsection{Sensitivity to GRB}

\begin{figure}
\begin{center}
\includegraphics[width=1.0\linewidth]{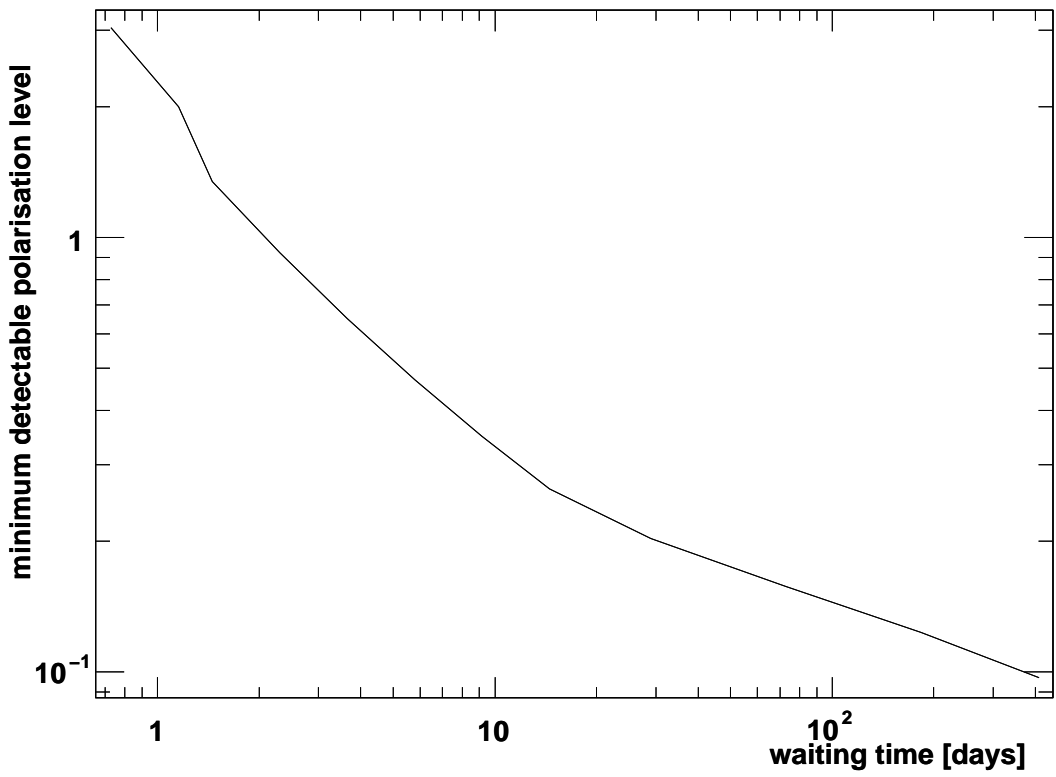}
\end{center}
\caption{Minimum detectable polarization level in function of waiting time}
\label{fig:wait}
\end{figure}

According to the GUSBAD catalog\cite{gusbad} there are 250 burst per year with a peak 
flux~$>$1 photon $cm^{-2}s^{-1}$ in the BATSE energy range 50 to 300 keV.
Extrapolation to our energy range (10 to 300 keV) gives a factor 2.5 bigger
flux.
Fig.~\ref{fig:wait} show the minimum detectable polarization
level in a GRB in function
of the mean waiting time in days for the occurrence of such a GRB according to
the GUSBAD catalog and taking into account that only $\frac{1}{3}$ events
are in our acceptance.

The duration of GRB is very variable. For short bursts which last less than 2 seconds,
the signal to background ratio is good enough even for low fluences. For intense 
long duration bursts, a polarization measurement in time slices of a few seconds is 
promising. Such a measurement is important since polarization could vary time.

\subsection{Sensitivity to the Crab nebula}
\begin{figure}
\begin{center}
\includegraphics[width=1.0\textwidth]{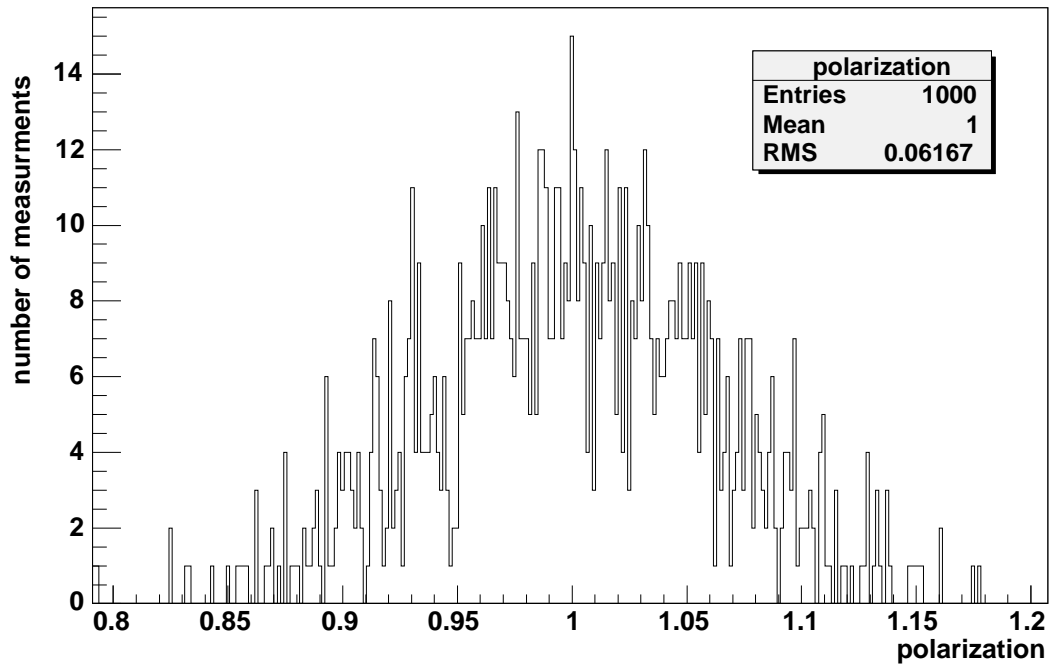}
\end{center}
\caption{Reconstructed degree of polarization for photons from the Crab
nebula, assuming 1000 measurements of 1000 seconds duration, assuming 100\%
polarized photons from the source.} 
\label{fig:crab}
\end{figure}

While the crab nebula is in the field of view of POLAR, a continuous polarization
measurement is feasible. In a 1 ks exposure, POLAR will collect about
56000 photons from the crab over a mean of 430000$\pm$655 photons of background. The 
expected results from 1000 such measurements are represented in Fig.~\ref{fig:crab}.
With the formula for the minimum detectable polarization level
we see that to detect a 
1\% polarization of the Crab we need to measure it for 180 ks.

\section{Conclusions}
Polarization is one the missing key to test different theoretical models of GRB.
We have presented the design of a compact (12 kg) detector to measure the polarization of GRB.
With that instrument we will know after one month of mission if highly polarized GRB exists.
In case GRB are not polarized, a limit down to 10\% polarization will be obtained after one year.
An array of such detector can be envisaged for increasing the polarization sensitivity
or the angular coverage.

\clearpage

\begin{table}
\begin{center}
\begin{tabular}{|c|c|c|c|c|c|}
\hline
process          & flux             &rate Hz      & PM hits Hz& readout rate Hz\\
\hline
strong GRB       & $20 cm^{-2}s^{-1}$      & $8 10^3$    & $4 10 ^3$  & $1.6 10^3$ \\
\hline
diffuse $\gamma$ & $2.46 cm^{-2}s^{-1} sr^{-1}$ & $1.8 10^4$  & $5 10^3$   & 430        \\
\hline 
cosmic           & $10^{-1} cm^{-2}s^{-1}sr^{-1}$ & 700        &  700       &   0       \\
\hline
proton $<$13MeV   & $10^3 cm^{-2}s^{-1}$    & $2.8 10^6$  & 0          &0          \\
\hline
proton $>$ 13MeV   & $10^2 cm^{-2}s^{-1}$    & $2.8 10^5$  & $2.8 10^5$ &0          \\
\hline
electron $<$ 500 keV &  $10^5 cm^{-2}s^{-1}$ & $2.8 10^8$  & 0          &0         \\
\hline
electron $>$ 500 keV & $3 10^3 cm^{-2}s^{-1}$ & $9 10^6$   & $ 5 10^6$  & 400 d + 50 r\\ 
\hline
\end{tabular}
\label{table:trigger}
\end{center}

\caption{The different processes contributing to signal and background and their impact on the
trigger rate. Flux is the theoretical flux. Rate is the flux integrated over
the fiducial surface. PM hits is the number of time a PM signal pass the lower threshold,
this incurred a 10 ns dead time. The
readout rate is number of time a coincidence in the correct energy window occur.}
\end{table}
\clearpage

\clearpage

\clearpage

\clearpage

\clearpage

\clearpage

\clearpage

\clearpage

\clearpage

\clearpage

\clearpage

\clearpage

\clearpage

\clearpage

\end{document}